\documentclass[a4paper,fleqn]{cas-dc}

\usepackage[authoryear,longnamesfirst]{natbib}
\usepackage{graphicx}
\usepackage{caption}
\DeclareUnicodeCharacter{22C6}{\ensuremath{\star}}
\def\tsc#1{\csdef{#1}{\textsc{\lowercase{#1}}\xspace}}
\tsc{WGM}
\tsc{QE}

\begin{document}
\let\WriteBookmarks\relax
\def\floatpagepagefraction{1}
\def\textpagefraction{.001}

\shorttitle{}    

\shortauthors{G. Montani, M. De Angelis, M.G. Dainotti}  

\title [mode = title]{Decay of dark energy into dark matter in a metric $f(R)$ gravity: effective running Hubble constant}  



%

\author[1]{Giovanni Montani}


\fnmark[1]

\ead{giovanni.montani@enea.it}


\credit{Conceptulisation and methodology}

\affiliation[1]{organization={ENEA, Nuclear Department, C. R. Frascati},
            addressline={Via E. Fermi 45}, 
            city={Frascati},
            postcode={00044}, 
            country={Italy}}
            
\affiliation[1]{organization={Physics Department, “Sapienza” University of Rome},
            addressline={P.le Aldo Moro 5}, 
            city={Roma},
            postcode={00185}, 
            country={Italy}}
            
\author[2]{Mariaveronica \ {De Angelis}}

\fnmark[2]

\cormark[1]
\ead{mdeangelis1@sheffield.ac.uk}


\credit{Methodology and Data Analysis}

\affiliation[2]{organization={School of Mathematical and Physical Sciences, University of Sheffield},
            addressline={Hounsfield Road}, 
            city={Sheffield},
            postcode={S3 7RH}, 
            state={South Yorkshire},
            country={United Kingdom}}

\author[3]{Maria Giovanna \ {Dainotti}}

\fnmark[3]

\ead{maria.dainotti@nao.ac.jp}


\credit{Data Analysis}

\affiliation[3]{organization={Division of Science, National Astronomical Observatory of Japan},
            addressline={ 2-21-1 Osawa, Mitaka}, 
            city={ Tokyo 181-8588},
            country={Japan}}
\affiliation[3]{organization={The Graduate University for Advanced Studies (SOKENDAI)},
            addressline={Shonankokusaimura, Hayama}, 
            city={ Miura District, Kanagawa 240-0115}}
\affiliation[3]{organization={Space Science Institute},
            addressline={Boulder, CO},
            country={USA}}
\cortext[1]{Corresponding author}



\begin{abstract}
We examine a modified late-Universe dynamics where dark energy decays into dark matter, within the framework of metric $f(R)$-gravity in the Jordan frame. After a detailed analysis of the modified $\Lambda \text{CDM}$ model, we introduce a theoretical diagnostic tool to capture the emergence of an effective running Hubble constant as a function of redshift. We then compare this theoretical model with the 40-bin analysis of the Supernova Pantheon sample. This comparison allows us to determine the value of the additional free parameter that appears in our model, beyond those of the standard $\Lambda \text{CDM}$ model. Our modified late Universe dynamics provides a good-quality fit to the binned data, improving upon the previous phenomenological interpretation based on a power-law decay. However, unlike the power-law model, our approach cannot be extrapolated to the recombination redshift to match the Hubble constant measured by the Planck satellite. In fact, the dynamics resulting from the binned Pantheon sample analysis address only weakly the Hubble tension between the SH0ES and the Planck Collaboration values of the Hubble constant. Here we
provide a convincing representation of the observed deviation of the
cosmological dynamics from the $\Lambda$CDM-one, as it out-stands from the low redshift observed sources.
\end{abstract}





\maketitle

\section{Introduction}

In recent years, the (average) 5$\sigma$ discrepancy between the Hubble constant ($H_0$) measurements by the SH0ES Collaboration (\cite{Brout:2022vxf} and references therein) and those inferred from the Planck satellite data \cite{Planck:2018vyg} has emerged as a central issue in modern cosmology, commonly referred to as the \lq \lq Hubble tension\rq \rq.

In the absence of a compelling astrophysical explanation -- such as a redshift evolution in the properties of Type Ia Supernovae (SNIae) when treated as standard candles (see e.g. \cite{Dainotti:2021vyp}) -- this tension is increasingly interpreted as potential evidence of new physics in the cosmological framework. This perspective is supported by several studies, including \cite{Schiavone:2022wvq,Montani:2023xpd,Montani:2024xys,Vagnozzi:2023nrq,PhysRevD.102.023518,PhysRevD.97.043528,DiValentino:2017oaw,DiValentino:2025sru}, and similar findings for tensions in the matter density parameter are discussed in \cite{PhysRevD.109.083511,Dainotti:2024aha}.

A significant step forward in understanding the nature of the Hubble tension has come from recent work by \cite{Dainotti:2021pqg,Dainotti:2022bzg,Dainotti:2025qxz}, which performed a binned analysis of SNIa data. These studies reveal a trend in which the inferred value of $H_0$ decreases as the redshift of the data bin increases. This result suggests that the tension is not merely a conflict between two distinct cosmological epochs -- \textit{i.e.} low-redshift SNIae ($z < 2$) and high-redshift CMB data ($z \sim 1100$) -- but instead reflects a continuous phenomenon manifesting within the SNIa data itself. This insight motivates a focus on late-time modifications to cosmic dynamics as a promising avenue to resolve the tension.

The observed slowing-down trend of $H_0$ across redshift bins was initially modeled in \cite{Dainotti:2021pqg} using a power-law form, $H_0 \propto (1+z)^{-\alpha}$, with $\alpha$ found to be of the order $\sim 10^{-2}$ (see also related results in \cite{Hu:2023jqc,Jia:2022ycc,Nojiri:2021dze}. In \cite{Dainotti:2022bzg}, a similar trend was reproduced by fitting each redshift bin with a $\Lambda$CDM model, allowing the matter density parameter to vary under a Gaussian prior rather than fixing it as in standard MCMC procedures. More recently, \cite{Dainotti:2025qxz} demonstrated that the same effective $H_0$ trend could be recovered using an extended sample, including a master collection of SNIae.

This power-law decay of the Hubble constant has been interpreted as the signature of a modified gravity framework, specifically within a $f(R)$-gravity theory formulated in the Jordan frame \cite{Schiavone:2022wvq,Schiavone:2024heb}. Alternative interpretations include quintessence models or evolving dark energy scenarios, as discussed in \cite{MONTANI2024101486,Montani:2024ntj}.

In this manuscript, we present a model of late-time cosmic evolution that combines two key mechanisms: i) an $f(R)$-gravity framework in the Jordan frame (see \cite{Sotiriou:2008rp,NOJIRI20171,Banerjee:2022ynv,NOJIRI2022115850}), and ii) a phenomenological process wherein dark energy decays into dark matter components. We demonstrate that both effects are essential to alleviating the Hubble tension. The required redshift-dependent scaling of $H_0$ -- bridging the SH0ES and Planck determinations -- is naturally achieved through the dynamics of a non-minimally coupled scalar field arising in modified gravity. Meanwhile, the inclusion of dark energy decay ensures this scalar mode has a positive mass, avoiding the tachyonic behavior it would otherwise exhibit.

Our model introduces only one new free parameter beyond those of the standard $\Lambda$CDM framework. This parameter is determined by fitting a theoretically derived expression for the running Hubble constant to binned SNIa data.

The combined theoretical construction and fitting analysis provide strong evidence for the model's efficacy in addressing the Hubble tension, offering a compelling scenario that integrates modified gravity and dark sector interactions.




\section{Dynamical scheme}

We consider a flat isotropic Universe, according to Planck data 
\cite{Lemos:2018smw} (for a different analysis see \cite{DiValentino:2019qzk}), which line element, in $c=1$ units, takes the form
\begin{equation}
	ds^2 = - dt^2 + a^2(t) dl^2
	\, , 
	\label{ba1}
\end{equation}
Here, $a(t)$ represents the cosmic scale factor governing the expansion of the Universe with respect to synchronous time $t$, while $dl^2$ denotes the standard infinitesimal Euclidean distance in three spatial dimensions.
We investigate the dynamics of the Universe within the framework of metric $f(R)$ gravity, formulated in the so-called Jordan frame \cite{Sotiriou:2008rp}, and assume the presence of the matter fluids that dominate the relevant phase of cosmic evolution.
The gravity-matter action takes the following form
\begin{equation}
	S_{g+m}=S_g+S_m \equiv 
	\frac{1}{2\chi}\int dt
a^3\left( \xi R - V(\xi )\right) + S_m
\, . 
	\label{ba2}
\end{equation}
Here, $\chi$ denotes the Einstein constant, and $S_m$ represents the action of the matter fields. All quantities depend solely on time, reflecting the assumption of spatial homogeneity. For simplicity, the unphysical fiducial volume in the action integral has been set to unity.
Specifically, the Ricci scalar $R(t)$ corresponding to the line element (\ref{ba1}) is given by $R = 6\dot{H} + 12H^2$, where $H(t) \equiv \dot{a}/a$ is the Hubble parameter, with the dot indicating differentiation with respect to time.
The scalar field $\xi(t)$ is non-minimally coupled to gravity, and its potential term $V(\xi)$ encapsulates the specific structure of the chosen $f(R)$ Lagrangian. These quantities are related through the expressions
\begin{equation} 
\xi = \frac{df}{dR}, \quad \quad V(\xi )\equiv R\frac{df}{dR} - f(R), 
\label{ba3} 
\end{equation} where we assume that $\frac{df}{dR}$ is an invertible function, allowing the Ricci scalar to be expressed as $R = R(\xi)$.
The first key dynamical equation arises from varying the action (\ref{ba2}) with respect to the scalar field $\xi$, noting that $\xi$ does not appear in the matter action $S_m$. This yields
\begin{equation} 
R = \frac{dV}{d\xi} = 6\dot{H} + 12H^2. 
\label{ba4} 
\end{equation} 
Since our cosmological model involves only two independent degrees of freedom -- $\xi(t)$ and the scale factor $a(t)$ (or equivalently the Hubble parameter $H(t)$) -- we require a second fundamental equation to fully determine the system. The energy density of the matter fluid can ultimately be expressed in terms of the scale factor, once the equation of state is specified.
Denoting the total energy density of the Universe by $\rho_{\text{tot}}(t)$, the second fundamental equation corresponds to the generalized Friedmann equation -- essentially, the Hamiltonian constraint arising from the time diffeomorphism invariance of the action (\ref{ba2}) (see \cite{book}). This leads to the relation 
\begin{equation}
	\xi H^2 = \frac{\chi}{3} \rho_{tot} - H \dot{\xi} + \frac{V(\xi )}{6}
	\, . 
	\label{ba5}
\end{equation}
established in \cite{Schiavone:2022wvq}.
\subsection{Specific assumptions}

We now focus on the late-time dynamics of the Universe, where the total energy density is assumed to comprise only two components: pressureless matter denoted by $\rho_m$, and dark energy denoted by $\rho_{de}$, whose pressure is taken to be exactly $P_{de}=-\rho_{de}$.
A key feature introduced in this framework is a phenomenological interaction between these two components, characterized by a constant rate $\bar{H} = \text{const.}$ For a detailed discussion in the context of a kinetic formulation, see \cite{Montani:2024pou}.
As a result, the coupled continuity equations governing the evolution of these two energy components are given by
\begin{align}
	\dot{\rho}_m + 3H\rho_m & = 
	\bar{H}\rho_{de},
	\label{ba6}\\
    	\dot{\rho}_{de} & = - \bar{H}\rho_{de}.
	\label{ba7}
\end{align}
It is important to note that $\rho_m \equiv \rho_{dm} + \rho_b$, where $\rho_{dm}$ refers to the cold dark matter component and $\rho_b$ to the baryonic matter. In our model, the energy transfer from dark energy is assumed to affect only the dark matter sector, while the baryonic component evolves independently according to the standard continuity equation $\dot{\rho}_b + 3H\rho_b = 0$.

If we interpret $\bar{H}$ as a positive constant, the coupled equations above can be viewed as describing a decay process in which dark energy is converted into dark matter. This interpretation serves as our reference scenario throughout the discussion. Nevertheless, more general frameworks could be introduced by including a second decay rate, representing a reverse energy flow from dark matter back into dark energy.

Since the total energy density of the Universe is naturally given by $\rho_{\text{tot}} \equiv \rho_m + \rho_{de}$, we can readily derive the following equation
\begin{equation}
	\dot{\rho}_{tot} + 3H\rho_{tot} = 3H \rho_{de},
    \label{ba8}
\end{equation}
where in the late Universe, the radiation contribution is negligible \cite{Kolb:1990vq}.
Following the approach discussed in \cite{Montani:2023xpd}, we choose those solutions that satisfy the following relation
\begin{equation}
	6H\dot{\xi} = V(\xi).
	\label{ba9}
\end{equation}
Imposing the condition above offers two main advantages: 
\begin{enumerate}[i)]
    \item It brings the Friedmann equation close to its standard form in General Relativity, with the only modification being the rescaling of the Einstein constant by $1/\xi$, as discussed in \cite{Montani:2024xys};
    \item The potential term $V(t) \equiv V(\xi(t))$ becomes a dynamical variable (this is accounted for by incorporating Equation (\ref{ba9}) into the dynamics). Furthermore, the form of $V(\xi)$ is dynamically determined a posteriori, once $\xi(t)$ and $V(t)$ are known, with the condition that $\xi(t)$ is an invertible function.
\end{enumerate}
Thus, the Friedmann equation \eqref{ba5} now rewrites as 
\begin{equation}
	H^2 = \frac{\chi}{3}
	\frac{\rho_{tot}}{\xi} . 
	\label{ba10}
\end{equation}
We see that our late Universe scenario is summarized by the closed system of ODEs \eqref{ba7}-\eqref{ba10} and \eqref{ba4}. In the following, we will examine the dynamical and physical predictions of this model, with particular focus on the potential alleviation of the Hubble tension.

\section{Dimensionless formulation}

In order to set up a dimensionless form of our late Universe modified dynamics, we first introduce a new time variable 
$x\equiv \ln (1+z)$, where $z\equiv a_0/a - 1$ ($a_0$ being the present-day value of the scale factor). 
It is immediate to recognise the validity of the following relation
\begin{equation}
	\dot{(...)} = -H(...)^{\prime} , 
	\label{ba11}
\end{equation}
where the prime denotes differentiation with respect to $x$. Furthermore, we introduce the following standard definitions
\begin{align}
	H_0 &\equiv H(x=0),\\ 
	\Omega_a &\equiv \frac{\chi \rho_a}{3H_0^2},  \\
	U &\equiv \frac{V}{6H_0^2},\\ 
	\label{ba12}
    \end{align}

where the subscript $a= \{\text{tot, matter, dark energy}\}$.
Hence, we can rewrite \eqref{ba4} and \eqref{ba7}-\eqref{ba10}. We start our analysis by observing that the ratio of \eqref{ba4} and \eqref{ba9}
results in 
\begin{equation}
	\frac{dU}{dx} = \left( 
	-2 + (\ln E)^{\prime}U\right),
\end{equation}
which admits the solution
\begin{equation}
     U = - \gamma 
	Ee^{-2x},
    \label{ba13}
\end{equation}
with $\gamma >0$ a constant. Then,
\begin{equation}
	\Omega_{de}^{\prime}  = 
	\frac{k}{E}\Omega_{de}, \qquad 
	k\equiv \frac{\bar{H}}{H_0},
	\label{ba14}
\end{equation}
where $\Omega_{de}(x=0)= 1-\Omega_m(x=0) \equiv 1-\Omega_m^0$, and
\begin{equation}
	\Omega_{tot}^{\prime} = 
	3\Omega_{tot} -3\Omega_{de},
	\label{ba15}
\end{equation}
where $\Omega_{tot}(x=0)=1$. Using now \eqref{ba13}, for the scalar field and the Friedmann equation we get
\begin{equation}
	\xi^{\prime} = \frac{\gamma}{Ee^{2x}}, \qquad \xi (x=0)=1,  
	\label{ba16}
\end{equation}
\begin{equation}
	E^2 \equiv \frac{H^2}{H_0^2} 
	= \frac{\Omega_{tot}}{\xi}, 
	\label{ba16b}
\end{equation}
respectively.

We observe that the proposed model includes four free parameters: $H_0$, $\Omega_m^0$, $k$, and $\gamma$, with the latter two being additional to the standard parameters of a $\Lambda$CDM model. In conclusion, we note that for the $f(R)$-gravity model to be viable, it is crucial that no tachyonic mode arises \cite{PhysRevD.72.083505,PhysRevLett.95.261102}. This condition is expressed as 
\begin{equation} 
\mu^2_{\xi}\equiv \frac{1}{3}\left( \xi \frac{d^2U}{d\xi^2} - \frac{dU}{d\xi}\right) \ge 0,
\label{ba18} 
\end{equation}
which must hold for the entire considered range of $x$.

\section{Study of the parameter space}

The SH0ES calibration of SNIa is based on the extrapolation of the luminosity-redshift relation up to $x = 0$ \cite{Brout:2022vxf}. Therefore, for a theoretical model to effectively address the Hubble tension, it must predict values for the deceleration parameter $q_0$ and the jerk parameter $j_0$ that are sufficiently close to those of the $\Lambda$CDM model \cite{Efstathiou:2021ocp}.
It is straightforward to verify that these parameters are defined as follows
\begin{align}
	q_0 &\equiv -1 + \frac{1}{2}
	(E^2)^{\prime}_{x=0},\\
	j_0 &\equiv 1 + \frac{1}{2}(E^2)^{\prime \prime}_{x=0} - \frac{3}{2}(E^2)^{\prime}_{x=0},
	\label{ba19}
\end{align}
Thus, for the $\Lambda$CDM-model, we get
\begin{align}
	q_0^{\Lambda \text{CDM}} &= -1+\frac{3}{2}\Omega_m^0,\\ j_0^{\Lambda \text{CDM}} &= 1.
	\label{ba20} 
    \end{align}
Now, by means of \eqref{ba14}-\eqref{ba16b}, we easily get
\begin{equation}
	q_0 = - 1 + \frac{1}{2}
	\left( 3\Omega_m^0 - \gamma\right).
	\label{21}
\end{equation}
To ensure that our model remains close to the corresponding $\Lambda$CDM value in \eqref{ba20}, we must impose the condition that $\gamma \ll 1$. Similarly, by neglecting higher-order terms in $\gamma^2$, we can derive the following expression for the jerk parameter (by appropriately differentiating the dynamical equations)
\begin{equation} 
j_0 \simeq 1 + \left(\frac{5}{2} - \frac{9}{4}\Omega_m^0\right) \gamma - \frac{3}{2}\left( 1 - \Omega_m^0\right) k. 
\label{ba22} 
\end{equation}
To reproduce $j_0 = 1$ as in the $\Lambda \text{CDM}$ model, we must require that the following relation for $k$ holds
\begin{equation} 
k = \frac{10-9\Omega_m^0}{6(1-\Omega_m^0)}\gamma.
\label{ba23} 
\end{equation}
Thus, $\gamma$ is the only additional parameter we deal with respect to a $\Lambda$CDM-model. 

We now limit the available range for the parameter $\gamma$ by requiring that the quantity $\mu^2_{\xi}$ be positive at least at $x = 0$. In this context, we can derive the following relation
\begin{equation}
\frac{dU}{d\xi} = \frac{1}{\xi^{\prime}}\frac{dU}{dx}.
\label{ba24} 
\end{equation}
Additionally, we observe that the following expression holds
\begin{align}
\mu^2_{\xi}(x=0) &= \frac{1}{3}\left[ \frac{\xi}{(\xi^{\prime})^2}\frac{d^2U}{dx^2} - \left( \frac{\xi\xi^{\prime\prime}}{(\xi^{\prime})^3} + \frac{1}{\xi^{\prime}}\right)\frac{dU}{dx}\right]_{x=0}\\ & = 2\Omega_m^0 +\frac{\Omega_m^0}{2\gamma}-\gamma +\gamma \frac{(10-9 \Omega_m^0)}{12(1-\Omega_m^0)}-\frac{8}{3}.
\label{ba25} 
\end{align}
From this, we see that to avoid a non-physical mode at $x = 0$ by considering $\Omega_m^0=0.298$ (see below), we must impose the condition $0 < \gamma < 0.071$. It is important to note that this condition is merely sufficient for the theory's viability. In fact, we must ensure that $\mu^2_{\xi} > 0$ for the entire $x$-interval (\textit{i.e.}, see Fig.\ \ref{fig:mass}).

To compare the $40$-bin analysis of the SNIa Pantheon sample data \cite{Dainotti:2021pqg,Dainotti:2022bzg}, which is discussed in the following section, we introduce the function $\mathcal{H}(z)$, as defined by \cite{Dainotti:2021pqg,Dainotti:2022bzg,Montani:2023xpd,Schiavone:2024heb}
\begin{equation}
	\mathcal{H}(z) \equiv 
	\frac{H(z)}{E_{\Lambda \text{CDM}}} = H_0\sqrt{\frac{\Omega_{tot}}{\xi\left[ \Omega_m^0(1+z)^3 + 1-\Omega_m^0\right]}},
	\label{ba26}
\end{equation}
where we made use of the redshift variable.
This function is simply an effective, redshift-dependent running Hubble constant and serves as an important diagnostic tool. To evolve $\mathcal{H}(z)$, we utilise \eqref{ba14}-\eqref{ba16b}, which are expressed in terms of the redshift using the relations $e^x = 1 + z$ and $(...)^{\prime} = (1 + z) \frac{d(...)}{dz}$.
\begin{figure}
    \centering
    \includegraphics[width=1\linewidth]{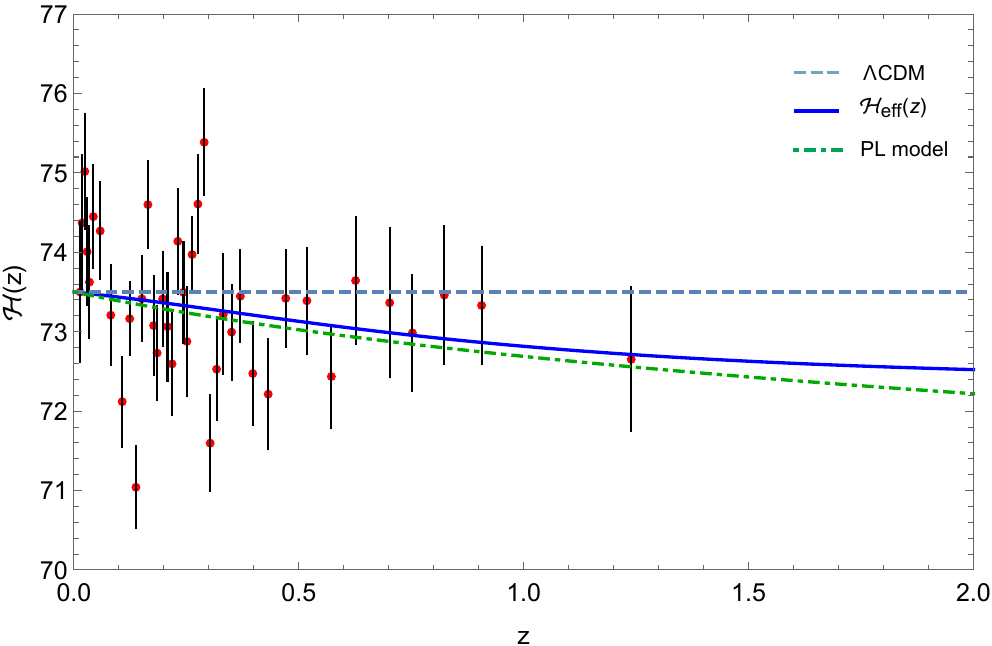}
    \caption{Plot of $\mathcal{H}_{\text{eff}}(z)$ from \eqref{ba26} (blue) for the best fit $\gamma$ in \eqref{eq: bestfit value gamma} and the fiducial values $H_0 = 73.5$ and $\Omega_{m0} = 0.298$. The green dot-dashed line represents the profile $H_0(1 + z)^{-0.016}$ (PL model). Red bullets are $H_0$ data \cite{Dainotti:2021pqg} with the corresponding error bars in $1\sigma$. We also depict the constant line $H_0 = 73.5$ for the base $\Lambda \text{CDM}$ model (light blue dashed line).}
    \label{fig:bestfit}
\end{figure}

\section{Data Analysis}
\begin{figure}
    \centering
    \includegraphics[width=1\linewidth]{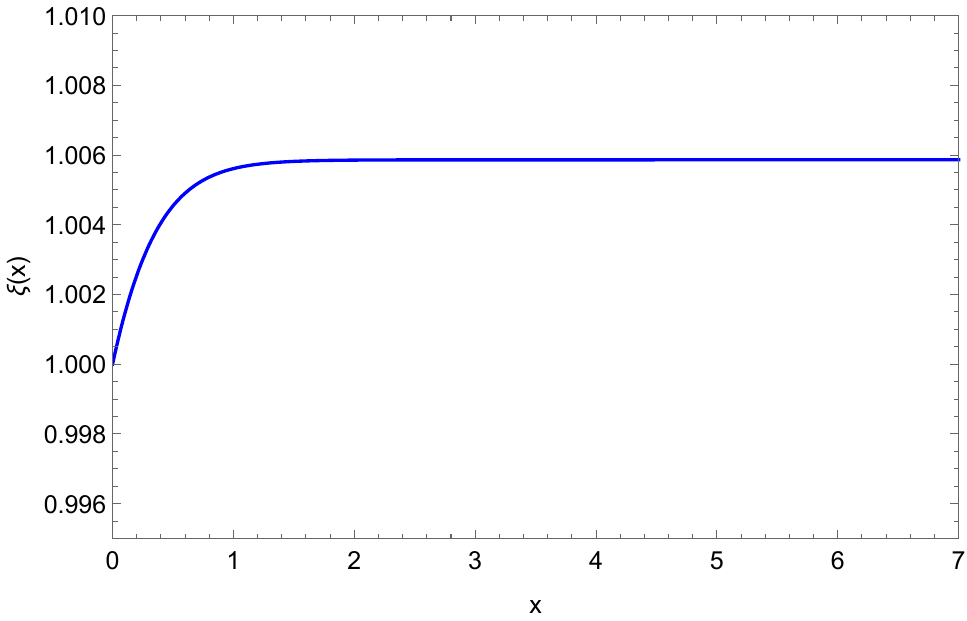}
    \caption{Behavior of the scalar field $\xi$ as a function of $x$ in the whole dynamical interval.}
    \label{fig:scalarfield}
\end{figure}
\begin{figure}
    \centering
    \includegraphics[width=1\linewidth]{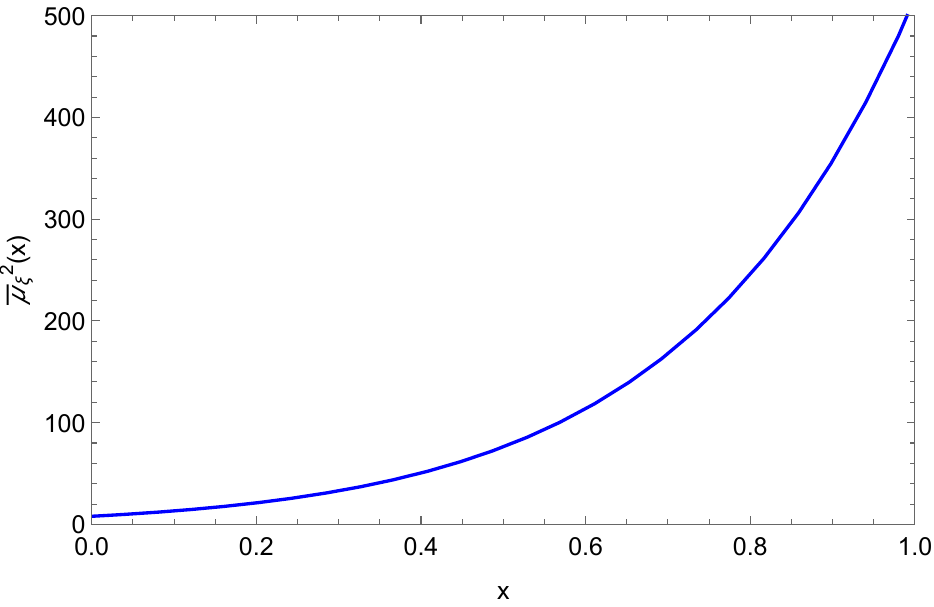}
    \caption{Plot of the normalised squared mass to $E^2(x)$ with $\gamma=0.0162$.}
    \label{fig:mass}
\end{figure}
Here, we present a comprehensive review of the methodology employed to conduct a binned analysis of the SN Pantheon sample \cite{Dainotti:2021pqg,Dainotti:2022bzg} and their fit with \eqref{ba26}. For clarity, the Hubble constant \(H_0\) is consistently expressed in units of km s\(^{-1}\) Mpc\(^{-1}\) throughout this study. 

The dataset under consideration is the Pantheon sample \cite{Pan-STARRS1:2017jku}, which compiles 1048 Type Ia supernovae from a variety of surveys. This sample is subdivided into 40 redshift bins, as illustrated along the \(x\)-axis in Fig.\ \ref{fig:bestfit}, with each bin containing an equal number of supernovae. In each bin, the redshift values are averaged, so that the bin centre represents the mean redshift of the supernovae it contains. Owing to the lower density of high-redshift supernovae, the central redshift of these bins is correspondingly biased towards lower values.

The observed distance modulus is defined as
\begin{equation}
\mu_{\mathrm{obs}} = m_B - M,
\end{equation}
where \(m_B\) is the apparent magnitude in the B band, corrected for both systematic and statistical uncertainties, and \(M\) denotes the absolute magnitude of a fiducial supernova, which includes corrections for colour and stretch. In accordance with previous studies, the distance moduli are averaged using the prescriptions provided in \cite{refId01,refId0}. The theoretical distance modulus is given by
\begin{equation}
\mu_{\mathrm{th}} = 5 \log_{10} d_L(z,H_0,\ldots) + 25,
\end{equation}
with \(d_L\) representing the luminosity distance, computed in megaparsecs within the framework of the base \(\Lambda\)CDM model. Corrections for the peculiar velocities of host galaxies are also incorporated.

To quantify the goodness-of-fit of our model to the data, we define the chi-squared statistic as
\[
\chi^2_{\mathrm{SN}} = \Delta\mu^T \cdot C^{-1} \cdot \Delta\mu,
\]
where
\[
\Delta\mu = \mu_{\mathrm{obs}} - \mu_{\mathrm{th}},
\]
and \(C\) is the \(1048 \times 1048\) covariance matrix provided in \cite{Pan-STARRS1:2017jku}. For each redshift bin, the value of \(H_0\) is estimated by varying \(H_0\) while keeping \(\Omega_{m0}\) fixed. The optimal fit in each bin is obtained via Markov Chain Monte Carlo (MCMC) techniques. It is noteworthy that the variation in \(\Omega_{m0}\) across the bins remains within \(2\sigma\) of its prior value \cite{Dainotti:2021pqg}.

The adoption of 40 bins reflects a deliberate compromise, ensuring that there are sufficient data points in each bin to allow for precise fitting, while simultaneously avoiding the introduction of excessive uncertainties that might arise from over-binning. Each bin contains approximately 26 supernovae, thereby providing an adequate data density for robust parameter estimation. Importantly, the declining trend observed in \(\mathcal{H}(z)\) is independent of the initial value chosen for \(H_0\), as the distance moduli in the Pantheon sample are presented with a fixed absolute magnitude. For instance, we adopt \(M = -19.245\), corresponding to \(H_0 = 73.5\) (see \cite{Dainotti:2022bzg}). An initial value for \(H_0\) is fixed in the first bin, from which \(M\) is determined as the sole free parameter; this value of \(M\) is then held constant in all subsequent bins, ensuring that the observed decrease in \(\mathcal{H}(z)\) is not an artefact of the fitting procedure.

Although the SH0ES collaboration reports a singular value of \(H_0\) from supernova data, recent analyses \cite{Dainotti:2021pqg,Dainotti:2022bzg} have demonstrated that a binned approach reveals a gradual decline in \(H_0\) with increasing redshift. If the Hubble tension indeed arises from new physics, such as an evolving dark energy component, one would expect to observe a continuous evolution of \(H_0\) across the redshift range, rather than a discrete jump between the local and cosmic microwave background (CMB) measurements.

We further stress that the function \(\mathcal{H}(z)\), as defined in \eqref{ba26}, is utilised as a diagnostic tool to determine whether the dataset is adequately described by the standard \(\Lambda\)CDM Hubble parameter, or if an alternative model (for example, one incorporating evolutionary dark energy) should be considered. In our analysis, \(\mathcal{H}(z)\) is expressed as a function of several free parameters, namely \(H_0\), \(\Omega_{m0}\), and \(\gamma\), all of which are, in principle, available for fitting. However, given the binned nature of our analysis, we adopt the fiducial values \(H_0 = 73.5\) and \(\Omega_{m0} = 0.298\) (as discussed previously), leaving \(\gamma\) as the sole free parameter.
For comparative purposes, we also consider the power-law profile
\[
\mathcal{H}(z) = H_0 (1+z)^{-\alpha},
\]
with the best-fit parameter \(\alpha = 0.016 \pm 0.009\) (hereafter referred to as the PL model) as determined in \cite{Dainotti:2021pqg}. 
A non-linear fit to the 40-bin distribution (see Fig.\ \ref{fig:bestfit}) of \(H_0\) yields the best-fit value for \(\gamma\) as
\begin{equation}
\gamma = 0.0162 \pm 0.0091,
\label{eq: bestfit value gamma}
\end{equation}
that lies in the range for which \eqref{ba18} is satisfied. From Fig. \ref{fig:scalarfield} we see that the scalar field increases towards a plateau and the
corresponding mass monotonically increases (see Fig.\ \ref{fig:mass}) from its positive value in $x=0$. This result confirms the viability of the obtained modified gravity model because it does not contain any tachyon mode for the whole system evolution.

\begin{figure}[h!]
    \centering
    \includegraphics[width=1\linewidth]{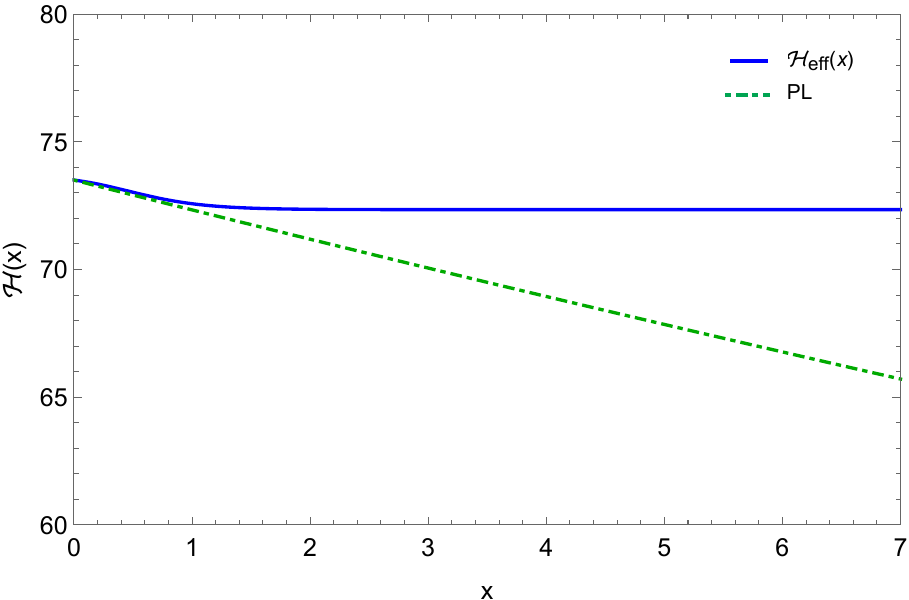}
    \caption{Profiles of $\mathcal{H}(x)$ corresponding to the proposed model (blue line) and the power-law behavior (green dot-dashed line).}
    \label{fig:Heff_x}
\end{figure}

We see that the value of $\gamma$, emerging from the non-linear fit of the Pantheon data, turns out to be too small to allow our model to account for the Hubble tension (despite, for a larger value, it could be possible).
Actually, the tension is only weakly attenuated and the value
of the effective running Hubble constant $\mathcal{H}(x)$ is,
asymptotically approaching the value $72.35$, larger than the one
predicted by the Planck data.
Thus, we are led to say that, although our fit of the Pantheon binned data is competitive with the one provided in \cite{Dainotti:2021pqg,Dainotti:2022bzg}, the latter has the pleasant property of being extrapolated to the value predicted by Planck within the errors, for the recombination value
$x\simeq 7$ (see Fig.\ \ref{fig:Heff_x}). In this respect, this phenomenological behavior remains the most interesting profile for the effective running Hubble tension with the redshift for the capability to incorporate
low and high $z$-value features of the cosmological dynamics.
Nevertheless, we have derived a dynamical representation that, so far, provides the best fit to the data from the binned description of the SNIa Pantheon sample. In this regard, it serves as a reliable and small modification of the $\Lambda \text{CDM}$ model, making it a viable candidate for SNIa physics.

\subsection{Statistical assessment of the fit}

To further assess the goodness of our fitting procedure, we compute the chi-squared statistic defined as
\begin{equation}
\chi^2 = \sum_{i=1}^{n} \frac{\left(y_i - f(x_i)\right)^2}{\sigma_i^2},
\end{equation}
where \( y_i \) are the observed values ($H_0$), \( f(x_i) \) are the model predictions, and \( \sigma_i \) the corresponding uncertainties. This allows for a direct evaluation of the discrepancy between the data and the theoretical model across the entire redshift range.
To account for the number of free parameters in the fit, we also calculate the reduced chi-squared statistic, given by
\begin{equation}
\chi^2_{\mathrm{red}} = \frac{\chi^2}{\nu},
\end{equation}
where \( \nu = n - k \) is the number of degrees of freedom, with \( k \) representing the number of fitted parameters. In our case, the model contains a single free parameter (\( k = 1 \)).

The results of the fit give
\begin{equation}
\chi^2 = 78.51, \quad \chi^2_{\mathrm{red}} = 2.01.
\end{equation}
While, for $\chi_{\text{red}}^{2 \text{(PL)}}=2.066$ and $\chi_{\text{red}}^{2 \text{($\Lambda \text{CDM}$)}}=2.176$. None of the models achieves a too-optimised fit, however, our model exhibits improved overall performance against the
standard $\Lambda \text{CDM}$ framework and PL model.  Furthermore, the $p$-value associated with the test is $p = 0.0001$,  indicating a statistically significant tension between the model and the data. Moreover, the coefficient of determination is rather low, $R^2 = 0.0468$, which suggests that although the model does not align well with the data, it also fails to capture much of the variance in the $H_0$ observations. This implies that the model may require refinement or the inclusion of additional degrees of freedom to better account for the structure of the data.

\begin{figure}
    \centering
    \includegraphics[width=1\linewidth]{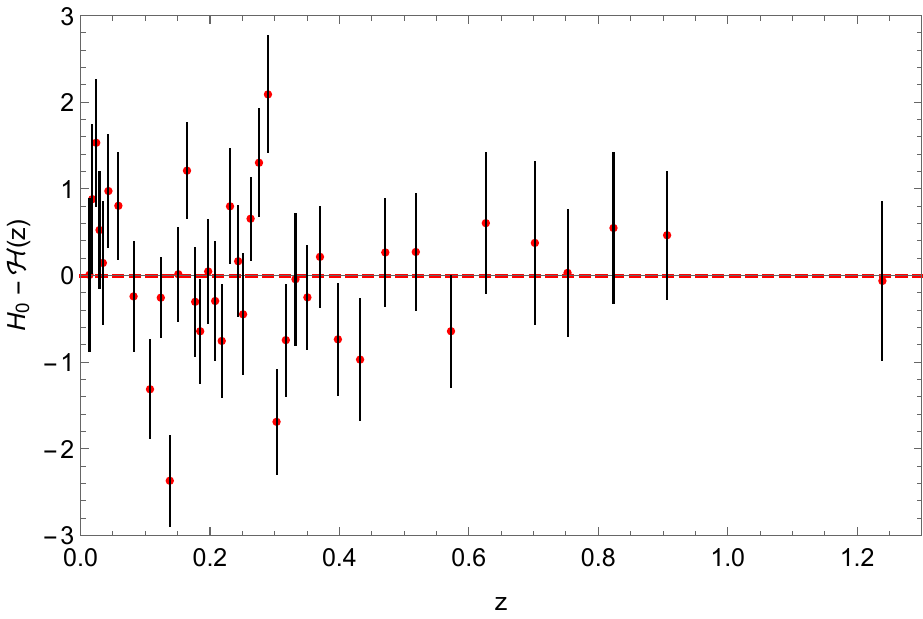}
    \caption{Plot of the residuals and respective errors. $H_0$ is the reference data value from \cite{Dainotti:2021pqg} and $\mathcal{H}_{\text{eff}}(z)$ is the model prediction (best fit). The red dashed line $y=0$ indicates the baseline where a perfect fit would result in all residuals being centered around zero.}
    \label{fig:res}
\end{figure}

Let the residuals be defined as
\[
\text{residuals} = y_i - f(x_i),
\]
where \( y_i \) are the observed values ($H_0$) and \( f(x_i) \) are the fitted model predictions $\mathcal{H}(z)$ (see Fig.\ \ref{fig:res}). We define the variance of residuals, assuming Gaussian errors
\[
\sigma^2 = \frac{1}{n} \sum_{i=1}^{n} (y_i - f(x_i))^2,
\]
where $n$ is the number of data points.
Hence, in order to calculate the Akaike Information Criterion (AIC) 
\[
\mathrm{AIC} = 2k - 2 \log \mathcal{L},
\]
and Bayesian Information Criterion (BIC)
\[
\mathrm{BIC} = k \log(n) - 2 \log \mathcal{L},
\]
we make use of the log-likelihood function defined as
\[
\log \mathcal{L} = -\frac{n}{2} \log(2\pi \sigma^2) - \frac{1}{2\sigma^2} \sum_{i=1}^{n} (y_i - f(x_i))^2,
\]
Therefore, using the best-fit values obtained from our model, we compute the residuals and subsequently evaluate the AIC and BIC. The analysis yields the following results
\begin{equation}
    \text{AIC}_{\text{eff}} = 102.73, \qquad \text{BIC}_{\text{eff}} = 104.42.
\end{equation}
These information criteria serve as quantitative tools to assess the relative performance of our model compared to standard cosmological scenarios. Lower values of AIC and BIC indicate a better trade-off between goodness of fit and model complexity.
For a comparison we also report $\text{AIC}_{\text{PL}}=103.12$, $\text{AIC}_{\Lambda \text{CDM}}=104.11$ and $\text{BIC}_{\text{PL}}=104.8$, $\text{BIC}_{\Lambda \text{CDM}}=104.11$. 
The Bayes factor between our model and the power-law scenario

\begin{align}
    \log \text{B}_{\text{eff/PL}} &= -\frac{1}{2}(\text{BIC}_{\text{eff}}-\text{BIC}_{\text{PL}})=0.19,\\
    B &\simeq 1.21,
    \end{align}
indicates a weak preference for this model.

\section{Concluding remarks}

We propose a modified $\Lambda$CDM framework that incorporates two conceptually distinct mechanisms. First, we adopt a metric $f(R)$-gravity formulation for a flat FLRW Universe and develop a dynamical representation in which the corresponding scalar potential emerges \emph{a posteriori}, as detailed in \cite{Montani:2023xpd,Montani:2024xys} Second, we introduce a phenomenological description of dark energy decay into dark matter, modelled through a transfer rate that governs the exchange of energy density between the two components \cite{Montani:2024pou}. 

The model is then analysed to ensure that its evolution near $z=0$ closely mimics that of the standard $\Lambda \text{CDM}$ scenario, thereby justifying the use of the luminosity–redshift relation for Type Ia supernovae. To this end, we examined the behavior of the deceleration parameter $q_0$, requiring that it matches the expression obtained in $\Lambda \text{CDM}$ in terms of the matter density parameter. The resulting dynamical framework introduces only one additional free parameter beyond the standard $\Lambda \text{CDM}$ model -- namely, the parameter $\gamma$, which governs the growth rate of the non-minimally coupled scalar degree of freedom arising in the metric $f(R)$ theory.
We tested the proposed dynamical framework against the 40-bin representation of the SNIa Pantheon sample \cite{Pan-STARRS1:2017jku}, which provides observational evidence for an effective Hubble constant that appears to decrease monotonically with redshift. This trend has been interpreted as a sign of tension between the observed expansion history and the standard cosmological model \cite{Schiavone:2024heb}, see also \cite{Montani:2024ntj}.
To this end, we compared the theoretical prediction of $\mathcal{H}(x)$, as determined by our model, with the data from the 40 SNIa bins. A non-linear fit was performed, keeping the values of $H_0$ and $\Omega_m^0$ fixed to those used in constructing the binned dataset, while allowing only the parameter $\gamma$ to vary in the fitting procedure.
Our analysis yields a best-fit value of $\gamma=0.016$, representing the most accurate match to the binned data obtained so far. Interestingly, the current model provides a better fit than the phenomenological power-law decay behavior initially introduced in \cite{Dainotti:2021pqg,Dainotti:2022bzg}, for which a theoretical derivation was offered in \cite{Schiavone:2022wvq} within a metric $f(R)$ gravity framework.
Despite the strong agreement between our model and the observed evolution of the effective Hubble parameter, the inferred value of $\gamma$ is too small to resolve the Hubble tension. The discrepancy is only mildly alleviated when extrapolating the behavior of $\mathcal{H}(x)$ up to the recombination epoch. In fact, we asymptotically reach the value of $72.35$, which is too large in comparison to the expected Hubble constant measured by the Planck Satellite \cite{Planck:2018vyg}, which is much larger than the observed one. At this point, the phenomenological power-law behavior appears more predictive and powerful, since it can reproduce a real Planck value for the Hubble constant, according to the measurements errors.

However, when focusing on the redshift range where the data are available -- the only relevant range for physical interpretations -- we find that the current model provides the best fit to the Pantheon sample binned data. Consequently, we observe that the combined dynamical effects of modified gravity and dark energy decay into dark matter present a viable scenario for describing the late-time evolution of the Universe, as inferred from the SNIa data.

\color{black}

\section*{Aknowledgment}
MDA is supported by an EPSRC studentship.

\printcredits

\bibliographystyle{cas-model2-names}

\bibliography{cas-refs}



\end{document}